\documentclass{jpsj-suppl}
\usepackage{txfonts} 
\usepackage{graphics}
\usepackage{amsfonts}
\usepackage{amsmath}
\usepackage{epsfig}
\usepackage{epsf}
\usepackage{graphicx}
\usepackage{dcolumn}
\usepackage{bm}
\usepackage{color}
\usepackage{array}

\title{Two-Photon-Exchange Effects and $\Delta(1232)$ Deformation}
\author{Hai-Qing \textsc{Zhou}$^{1}$ and Shin Nan \textsc{Yang}$^{2}$}

\inst{$^{1}$ Department of Physics, Southeast University, NanJing 211189, China \\
$^{2}$Department of Physics and Center for Theoretical Sciences, National Taiwan University, Taipei 10617, Taiwan}

\email{$^1$zhouhq@seu.edu.cn, $^2$snyang@phys.ntu.edu.tw}

\recdate{Oct 16, 2016}

\abst{The two-photon-exchange (TPE) contribution in   $ep\rightarrow ep\pi ^0$  with $W=M_{\Delta}$ and small $Q^2$ is calculated and its corrections to the ratios of electromagnetic transition form factors $R_{EM} = E_{1+}^{(3/2)}/M_{1+}^{(3/2)} $ and $R_{SM} = S_{1+}^{(3/2)}/M_{1+}^{(3/2)}$,   are analysed.  A simple hadronic model is used  to estimate the TPE amplitude.   Two phenomenological models, MAID2007 and SAID, are used to approximate the full $ep\rightarrow ep\pi ^0$ cross sections which contain both the TPE and the  one-photon-exchange (OPE) contributions.  The genuine the OPE amplitude is then extracted from an integral equation by iteration. We find that the TPE contribution is not sensitive to   whether MAID or SAID is used as input in the region with  $Q^2<2$ GeV$^2$.
 It gives small correction to $R_{EM}$  while for $R_{SM}$, the correction is  about  -10\% at small $\epsilon$ and about $1\%$ at large $\epsilon$  for $Q^2\approx2.5$ GeV$^2$. The large correction from TPE at small $\epsilon$ must be included in the analysis to get a  reliable extraction of  $R_{SM}$.}

\kword{two-photon-exchange, pion production, $\Delta(1232)$ deformation}

\begin{document}
\maketitle


The study of the structure of the hadrons is one of the most important way to understand the non-perturbative properties of QCD.
As the first excited  state of the nucleon,  the  structure of $\Delta(1232)$ plays a special role  as it has been established that $\Delta$ deforms.  It provides a non-trivial test for  theoretical models \cite{Pascal07}. The electromagnetic excitation of the $\Delta(1232)$   provides a way to determine  the deformation by measurement of the electric E2 and Coulomb C2 transitions in the electromagnetical (EM) production of pion in the $\Delta(1232)$ region. Recently, a few precise measurements of the multipoles $M_{1^+},E_{1^+}$ and $S_{1^+}$ related to this transition have been performed \cite{ep-eppi0-Ex-data}. The precision of the measurements of the cross sections is  close to $1\%$ which  implies that electromagnetic radiative corrections should be included in the theoretical analysis  for the extraction to be reliable.

It has been established that the two-photon-exchange (TPE) effects contribute non-negligibly to the elastic $ep$ scatterings \cite{Arrington11} such that  the TPE  effects  must be taken into account in the  extraction of the ratio of the nucleon form factors $G_E/G_M$ reliably. It is  natural to ask how   the TPE processes would contribute to the   $\Delta(1232)$ excitation when  aiming at   precise determination  of $M_{1^+}, E_{1^+}$ and $S_{1^+}$.   The  TPE corrections to the EM excitation of $\Delta(1232)$ at high $Q^2$ have been considered  in  \cite{Pascalutsa-2006} in partonic approach with the use of GPDs. However,   TPE contributions at low $Q^2$ where the deformation of the $\Delta(1232)$ is inferred, have not been estimated.  In addition, in  \cite{Pascalutsa-2006}, only the $\Delta(1232)$ pole contribution, but not the Born term, is considered for the one-photon-exchange (OPE) mechanism in calculating the interference between OPE and TPE amplitudes. However, it is well-known that the Born term plays important role within the OPE framework for pion EM production \cite{Pascal07}. In this work, we estimate the TPE corrections to $M_{1^+},E_{1^+}$ and $S_{1^+}$ in the EM excitation of $\Delta(1232)$ at low momentum transfer $Q^2$  with the inclusion of the Born term in the OPE amplitude.

Within OPE approximation, the unpolarized cross section of the $ep\rightarrow ep\pi ^0$ can be expressed as
\begin{eqnarray}\label{eq:OPE-sigma}
& &\frac{d^5\sigma^{1\gamma}}{d\Omega_f dE_f d\Omega}  =  C|\mathcal{M}^{1\gamma}\mathcal{M}^{1\gamma*}| \equiv  \Gamma \{\sigma_0^{1\gamma}+\sqrt{2\epsilon(1+\epsilon)}\sigma_{LT}^{1\gamma}\cos\phi + \epsilon \sigma_{TT}^{1\gamma} \cos 2\phi \}, \label{OPEunpolCrS}
\end{eqnarray}
where $\sigma_0^{1\gamma}=\sigma_T^{1\gamma} + \epsilon
\sigma_L^{1\gamma}$, $\phi$ is the pion center of mass azimuthal angle with
respect to the electron scattering plane,  $\epsilon$ is the transverse polarization of the virtual photon,
$\Gamma$ is the virtual photon flux, and $C$ is a kinematical constant.
The virtual photon differential cross sections
($\sigma_{T}^{1\gamma},\sigma_{L}^{1\gamma},\sigma_{LT}^{1\gamma},\sigma_{TT}^{1\gamma}$)
are all functions of the center of mass energy $W$,  four momentum transfer squared $Q^2$,
and the pion center of mass polar angle $\theta_{\pi}$ measured
from the momentum transfer direction. When using the multipoles as  inputs, the cross sections can be expressed as the functions of multipoles explicitly as $\sigma_{0,LT,TT}^{th}(Z_l,\theta_\pi)$,
where $Z_l$'s denote the multipoles $E_{l^\pm},M_{l^\pm}$ and $S_{l^\pm}$. The superscript "$th$" is meant to emphasize that the expression $\sigma_{0,LT,TT}^{th}(Z_l,\theta_\pi)$ is fixed from theoretical consideration.

With the TPE contributions included, the unpolarized cross sections can be  expressed as, in the same form as the  OPE case of Eq. (\ref{OPEunpolCrS}) \cite{Pascalutsa-2006},
\begin{eqnarray}\label{eq:OPE-sigma}
 \frac{d^5\sigma^{1\gamma+2\gamma}}{d\Omega_f dE_f d\Omega} &\simeq& C\{ |\mathcal{M}^{1\gamma}\mathcal{M}^{1\gamma^*}|+2Re[\mathcal{M}^{2\gamma}\mathcal{M}^{1\gamma*}]\}\nonumber\\
&\equiv&  \Gamma \{\sigma_0^{1\gamma+2\gamma}+\sqrt{2\epsilon(1+\epsilon)}\sigma_{LT}^{1\gamma+2\gamma}\cos\phi + \epsilon \sigma_{TT}^{1\gamma+2\gamma} \cos 2\phi \},
\end{eqnarray}
where  the cross sections $\sigma_{T,L,LT,TT}^{1\gamma+2\gamma}$'s are now  $\epsilon$ dependent.
 Within the OPE plus TPE
approximation, $d\sigma^{1\gamma+2\gamma}/d\Omega_\pi$ would be what is measured experimentally and we may hence write
\begin{align}
& \frac{d\sigma^{ex}}{d\Omega_\pi} \simeq C\{|\mathcal{M}^{1\gamma}(X^{1\gamma}_{1^\pm},\bar{Z}^{1\gamma}_{l^\pm})|^2+ 2Re[\mathcal{M}^{1\gamma *}(X^{1\gamma}_{1^\pm},\bar{Z}^{1\gamma}_{l^\pm})\mathcal{M}^{2\gamma}]\},
\label{multipoles-TPE}
 \end{align}
where  $X^{1\gamma}_{1^+}=(E^{1\gamma}_{1^+}, M^{1\gamma}_{1^+}, S^{1\gamma}_{1^+})$ denotes the multipoles pertaining to the  $\Delta$ excitation channel, while $\bar{Z}^{1\gamma}_{l^\pm}$ represent all other multipoles. The superscript $1\gamma$ means that they are the genuine multipoles extracted from the data with TPE effects removed.   The functional dependence
of $\mathcal{M}^{1\gamma}$ on multipoles $(X_{1^\pm}$ and $\bar{Z}_{l^\pm})$ are well known.

\begin{figure}[htbp]
\center{\epsfxsize 1.8 truein\epsfbox{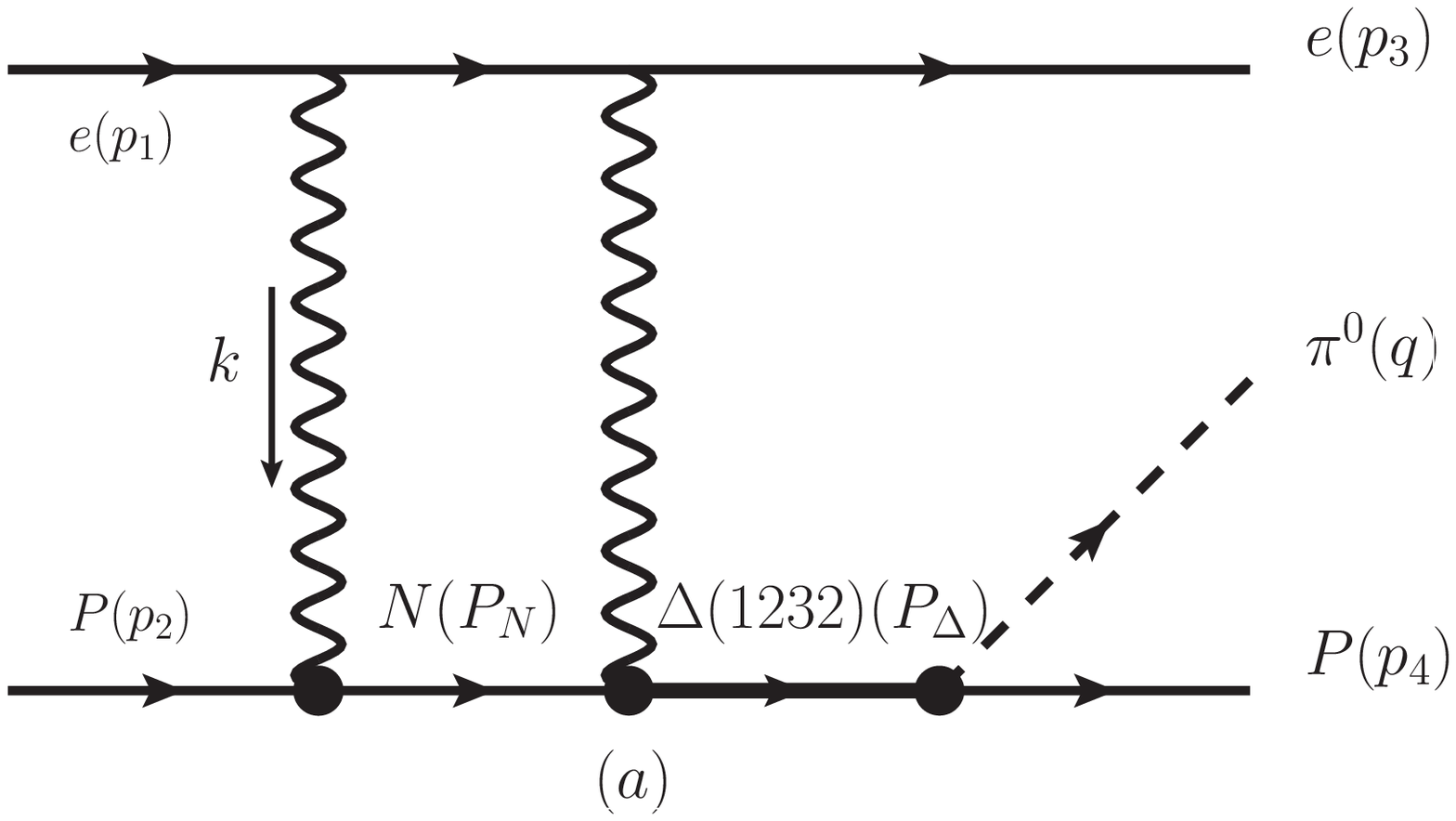}\epsfxsize 1.6 truein\epsfbox{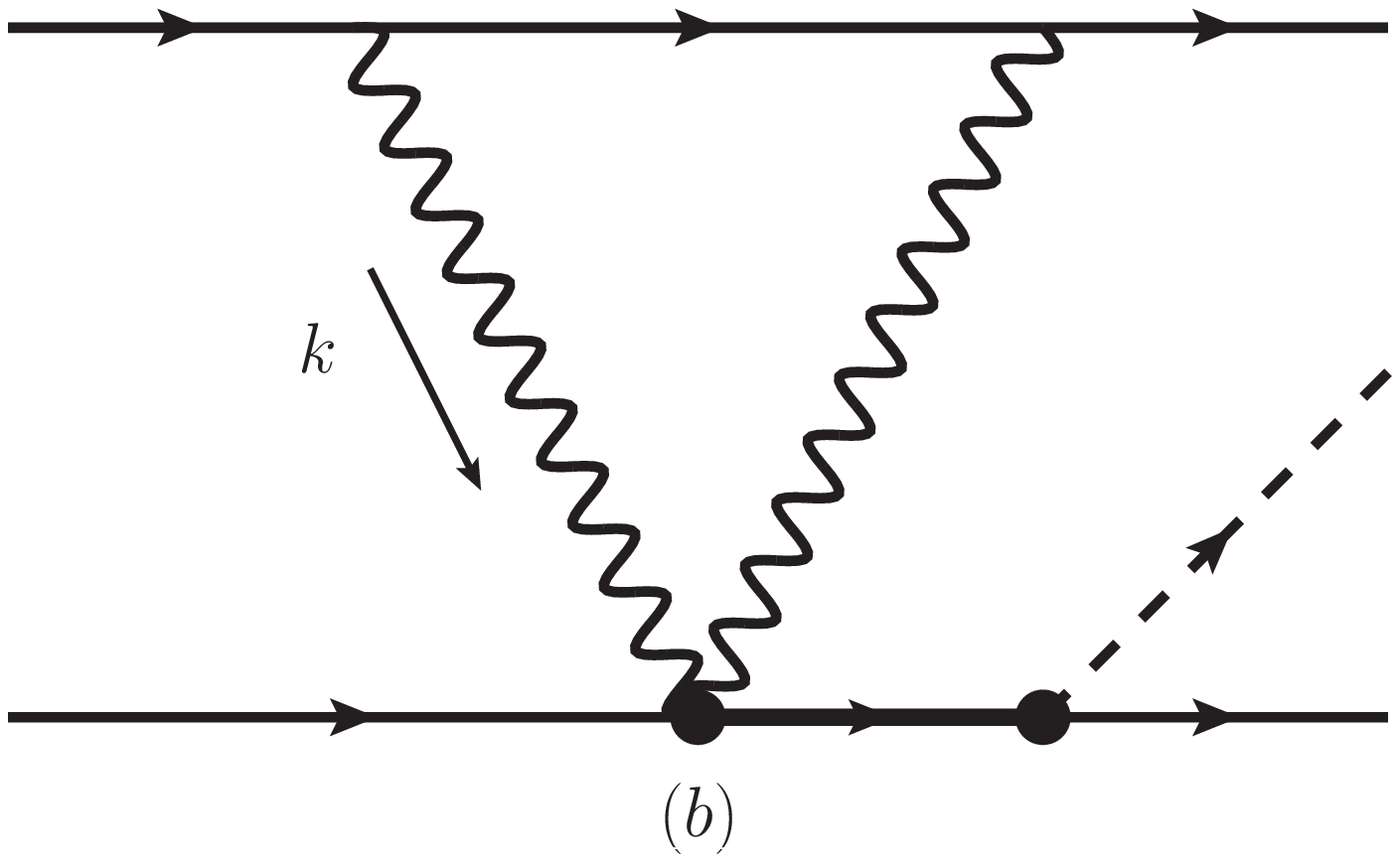}}
\caption{TPE diagrams for $ep \rightarrow e\Delta \rightarrow ep\pi ^0$, only the box and contact diagrams are showed, the x-box diagram is  not showed.}
\label{figure:ep-eppi0-TPE}
\end{figure}

The TPE diagrams we consider are depicted in Fig. \ref{figure:ep-eppi0-TPE}. The crossed box diagram is not shown while the contact diagram of Fig. \ref{figure:ep-eppi0-TPE}(b) is required by the gauge invariance \cite{Kondratyuk-2006-NPA}. As in \cite{Zhou07}, only the elastic intermediate states are taken into account in this study. Since we are interested in the TPE effects on the $\Delta$ peak $W=M_\Delta$, only the diagrams with $\Delta$ excited by the TPE are considered. In the evaluation of the diagrams of Fig. \ref{figure:ep-eppi0-TPE},
     vertices $\gamma^*NN$, $\gamma^*N\Delta$, $\gamma^*\gamma^*N\Delta$, and the $\Delta$ propagator  are taken from \cite{Kondratyuk-2006-NPA} and\cite{Zhou15}.  The gradient coupling is
 used for  $\pi^0 N \Delta$ vertex. The simple hadronic model as employed in \cite{Zhou15} is followed to evaluate diagrams of Fig. \ref{figure:ep-eppi0-TPE} for the amplitude $\mathcal{M}^{2\gamma}$.

To obtain the genuine OPE multipoles $Z^{1\gamma}_{l^\pm}$ from Eq. (\ref{multipoles-TPE}), we use the cross sections given by the two phenomenological models (MAID2007 \cite{MAID2007} and SAID \cite{SAID}) to approximate $\frac{d\sigma^{ex}}{d\Omega_\pi}$. In addition, $\bar{Z}^{1\gamma}_{l^\pm}$ is taken fixed such that only the three multipoles $X_{1^+}=(E_{1^+}, M_{1^+}, S_{1^+})$  remained to be determined. Furthermore, the phases of $E_{1^+}^{1\gamma}, M_{1^+}^{1\gamma}$ and $S_{1^+}^{1\gamma}$ are fixed to be the same as the respective input model as required by  unitarity. This reduces the variables
in Eq. (\ref{multipoles-TPE}) to only three numbers, namely, $|E_{1^+}|$, $|M_{1^+}|$, and $|S_{1^+}|$.
The nonlinear equation of Eq. (\ref{multipoles-TPE}) can solved via iteration.  The multipoles $X_{1^+}$ given by either MAID2007 (with $l\leq 8$) or SAID (with $l\leq 5$), depending on which model is used to approximate the experimental cross sections, are used for $M^{1\gamma}$ in Eq. (\ref{multipoles-TPE}), in the first iteration. We find that only one iteration is sufficient.

In the first iteration, we have the followings from Eq. (\ref{multipoles-TPE}),
\begin{align}
& \frac{d\sigma^{1,M/S}}{d\Omega_\pi}\equiv \{
\frac{d\sigma^{ex,M/S}}{d\Omega_\pi} -  2CRe|\mathcal{M}^{1\gamma *}(X^{0,M/S}_{1^\pm},\bar{Z}^{0,M/S}_{l^\pm})\mathcal{M}^{2\gamma}|\}=C|\mathcal{M}^{1\gamma}(X^{1,M/S}_{1^\pm},\bar{Z}^{0,M/S}_{l^\pm})|^2,
\label{multipoles-TPE-1st}
 \end{align}
where superscript $M/S$   refers to either MAID or SAID is used to approximate the cross sections and the corresponding multipoles. Superscripts $0$ and $1$ denotes the initial guess and the
resulting multipoles obtained from first iteration,  e.g., $X^{0,M}_{1^\pm},\bar{Z}^{0,M}_{l^\pm}=X_{1^\pm}(MAID), \bar{Z}_{l^\pm}(MAID)$.

The problem now is   how to determine the genuine OPE multipoles $X^{1,M/S}_{1^\pm}$'s from Eq. (\ref{multipoles-TPE-1st}) with TPE contributions removed. This can be done by first noting that the multipole dependence
of $\sigma^{1,M/S}$ of Eq. (\ref{multipoles-TPE-1st}) should be the same as $\sigma^{1\gamma}$ of Eq.  (\ref{eq:OPE-sigma}), which we define as $\sigma_{0,LT,TT}^{th}(Z_l,\theta_\pi)$ earlier.
We then determine the absolute values of $M_{1^+}^{1\gamma},E_{1^+}^{1\gamma}$ and $S_{1^+}^{1\gamma}$ by minimizing the following  $\chi^2$
\begin{eqnarray}\label{eq:2}
\chi^2 \equiv \sum\limits_{j=0, LT, TT}\sum\limits_{\theta_i = 1^o}^{179^o} \omega_j\{\sigma_{j}^{1,M/S}(
\theta_i)-\sigma_{j}^{th}(M_{1^+}^{1,M/S},E_{1^+}^{1M/S}, S_{1^+}^{1,M/S},\theta_i)\}^2
\end{eqnarray}
where  $\omega_j$'s are the weights of cross sections $\sigma_{0, LT, TT}$ used in the fitting. We choose equal weights $\omega_j=1$ in this preliminary study.
\begin{figure}[h]
\center{\includegraphics[scale=0.5]{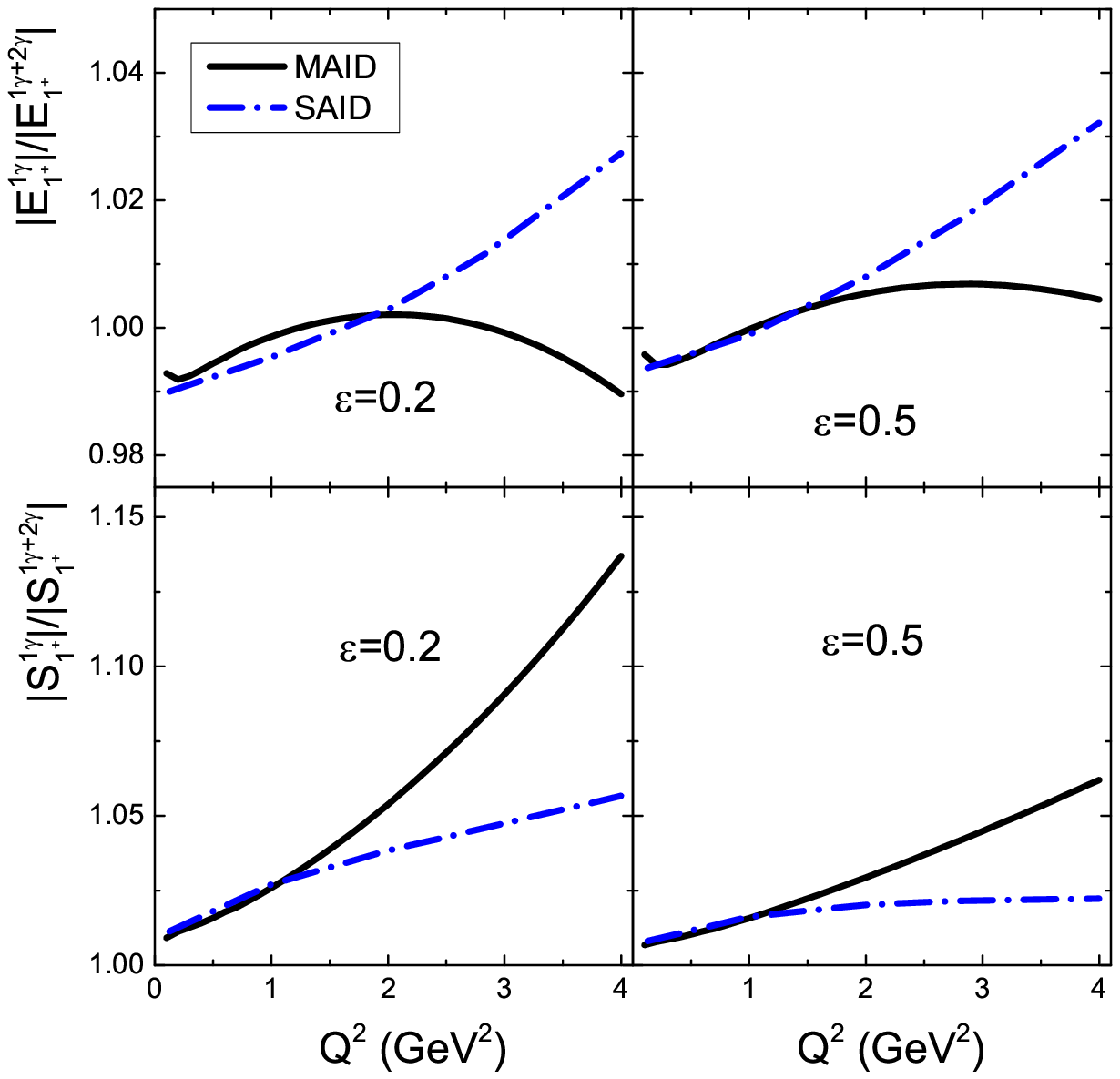}\includegraphics[scale=0.5]{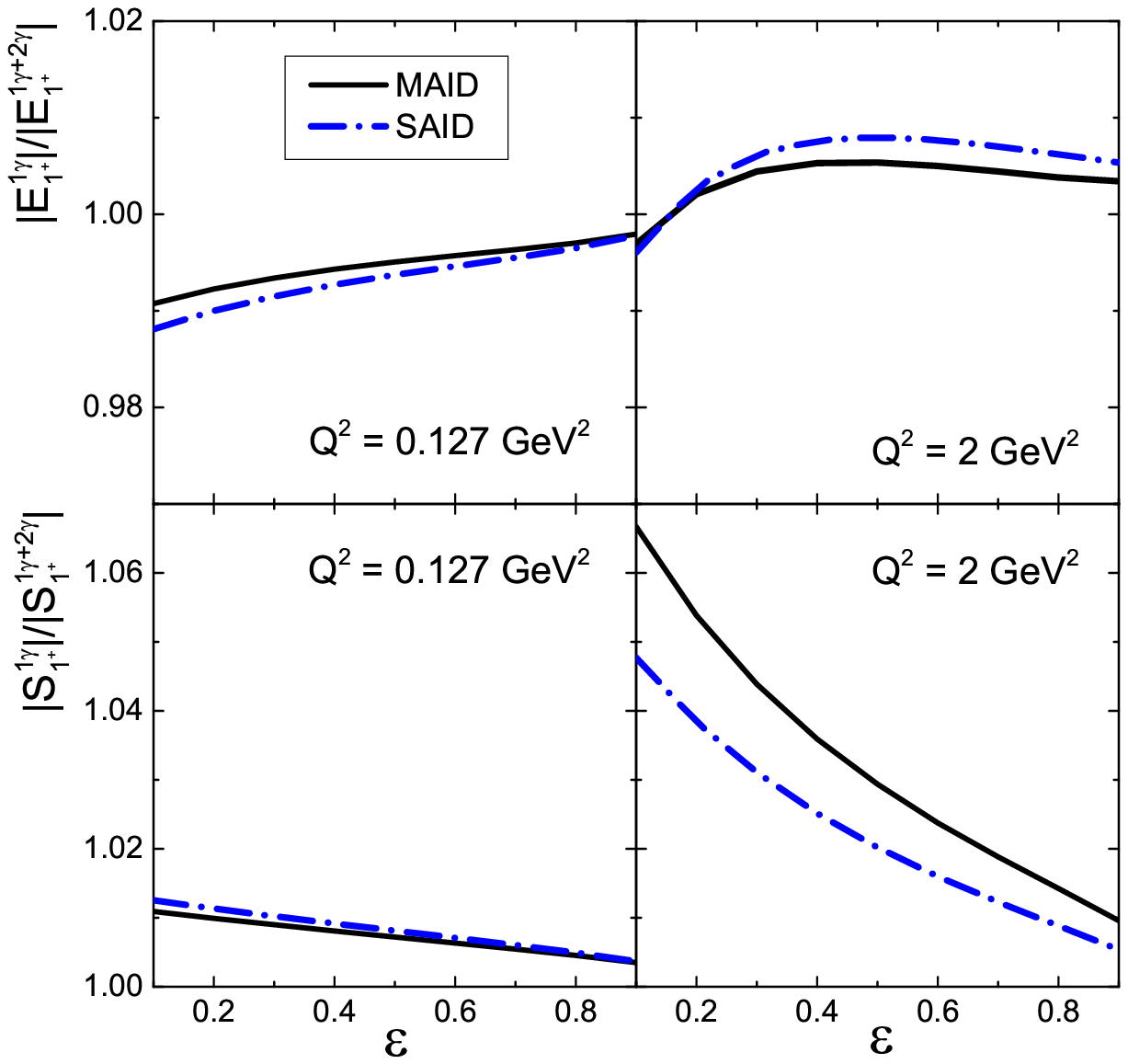}}
\caption{TPE corrections to the extracted multipoles $E_{1^+}$ and $S_{1^+}$ with $\omega_j=1$. The solid curves refer to the results with MAID2007 as input, and the dashed curves  refer to the results with SAID as input.} 
\label{figure:E1S1-TPE-e}
\end{figure}


We find that the TPE corrections to $M_{1^+}$ are very small. The TPE corrections to the mulitoples $E_{1^+}$ and $S_{1^+}$ with the MAID2007 and SAID as inputs are presented in the Fig.  \ref{figure:E1S1-TPE-e}, where the ratio
between the multipoles $X_{1^+}^{0,M/S}$ used in the input and the extracted genuine multipoles $X_{1^+}^{1\gamma,M/S}$ after TPE effects are removed. Note that the multipoles $X_{1^+}^{0,M/S}$'s which contain some TPE effects are
labelled as $X_{1^+}^{1\gamma + 2\gamma}$ in the figures. It is seen that the $\epsilon$  dependence of the TPE corrections  are similar with different models

\begin{figure}[tbh]
\begin{minipage}{0.50\linewidth}
as inputs, while the  $Q^2$ dependence of the corrections are rather different for $Q^2>2$ GeV$^2$. In this higher $Q^2$ region,   the SAID multipoles agree better with the experiments.

~~The absolute TPE corrections to the $R_{EM}$ and $R_{SM}$ at fixed $\epsilon$ are presented in the Fig. 3  where $R_{EM}\equiv  Im[E_{1^+}]/Im[M_{1^+}]$ and $R_{SM}\equiv Im[S_{1^+}]/Im[M_{1^+}]$. It is
interesting to note that even though the obtained TPE corrections $S_{1^+}^{1\gamma}/S_{1^+}^{1\gamma+2\gamma}$ are very different with different input models in the $Q^2\sim 2-4 $GeV$^2$ region as
seen in Fig. 2, the corrections $\delta R_{SM}$ are almost same. When comparing our results with those given in  \cite{Pascalutsa-2006}, we see that when $\epsilon=0.2$ our $\delta R_{EM}$ are much smaller than theirs, while our $\delta R_{SM}$ are much larger than theirs in the intermediate $Q^2$. One of the main reason of this difference lies in the fact that only the $\Delta$ pole term is included  for $\mathcal{M}_{1\gamma}$ in \cite{Pascalutsa-2006}. Namely,
the background contributions to $\mathcal{M}_{1\gamma}$ are not considered there when calculating the interference term between $\mathcal{M}^{1\gamma}$ and $\mathcal{M}^{2\gamma}$ .
\end{minipage}
\hfill
\begin{minipage}{0.46\linewidth}
\center{\includegraphics[scale=0.55]{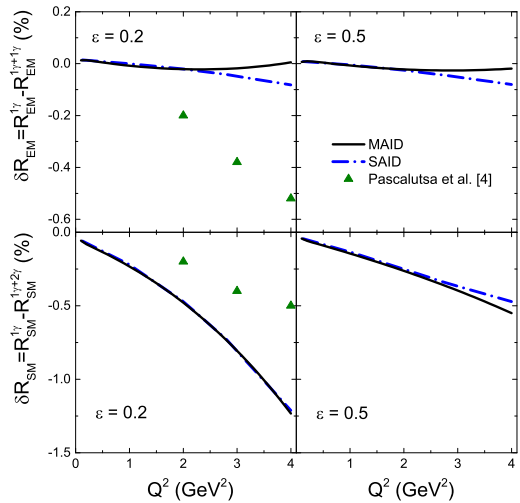}}
\caption{ TPE corrections $\delta R_{EM,SM}\equiv R_{EM,SM}^{1\gamma}-R_{EM,SM}^{1\gamma+2\gamma}$ with $\omega_j=1$. Notations same as Fig. \ref{figure:E1S1-TPE-e}.}
\end{minipage}
\label{figure:REMRSM-TPE-QQ}
\end{figure}

In summary, we evaluate the TPE corrections in the $ep \rightarrow ep \pi^0$ at $W=M_\Delta$ in the low $Q^2$ region in the hadronic approach. We include  the background contribution in the OPE amplitude.  We find that TPE corrections $\delta R_{SM}=R_{SM}^{1\gamma}-R_{SM}^{1\gamma+2\gamma}$ are not sensitive w.r.t. whether MAID2007 or SAID  is used as input model. Our results   differ  considerably with   \cite{Pascalutsa-2006} in $Q^2\sim 2-4$ GeV$^2$
region.\\

\noindent ACKNOWLEDGEMENTS\\
\\
This work is supported in part
by the National Natural Science Foundations of China under Grant No.
11375044, the Fundamental Research Funds for the Central Universities under Grant No. 2242014R30012 for H.Q.Z. and the National Science Council of the Republic of China (Taiwan) for S.N.Y. under grant No. NSC101-2112-M002-025.
H.Q.Z. would also like to gratefully acknowledge the support of
the National Center for Theoretical Science (North) of the National
Science Council of the Republic of China for his visit in the January of 2016.

\end{document}